\documentclass{ieeeaccess}
\usepackage{cite}
\usepackage{amsmath,amssymb,amsfonts}
\usepackage{algorithmic}
\usepackage{graphicx}
\usepackage{textcomp}
\usepackage{color,soul}
\def\BibTeX{{\rm B\kern-.05em{\sc i\kern-.025em b}\kern-.08em
    T\kern-.1667em\lower.7ex\hbox{E}\kern-.125emX}}
\begin{document}
\history{Date of publication xxxx 00, 0000, date of current version xxxx 00, 0000.}
\doi{10.1109/ACCESS.2017.DOI}

\title{Integrating Statistical and Machine Learning Approaches for Neural Classification}
\author{\uppercase{Mehrad Sarmashghi}\authorrefmark{1}, 
\uppercase{Shantanu P. Jadhav\authorrefmark{3}, and Uri T. Eden}.\authorrefmark{2},
}
\address[1]{Division of Systems Engineering, Boston University, Boston, MA 02215 USA (e-mail: pmsarmashghi@gmail.com)}
\address[3]{Department of Psychology, Brandeis University, Waltham, MA 02453 USA (e-mail: shantanu@brandeis.edu)}
\address[2]{Department of Mathematics and Statistics, Boston University, Boston, MA 02215 USA (e-mail: tzvi@bu.edu).}
\tfootnote{"This work was supported by Simons Foundation Grant 542971 and National Institutes of Health (NIH) Grants R01-
MH105174, and RO1-MH112662."}

\markboth
{Author \headeretal: Preparation of Papers for IEEE TRANSACTIONS and JOURNALS}
{Author \headeretal: Preparation of Papers for IEEE TRANSACTIONS and JOURNALS}

\corresp{Corresponding author: Mehrad Sarmashghi (e-mail: pmsarmashghi@gmail.com).}
.

\begin{abstract}
Neurons can code for multiple variables simultaneously and neuroscientists are often interested in classifying neurons based on their receptive field properties. Statistical models provide powerful tools for determining the factors influencing neural spiking activity and classifying individual neurons. However, as neural recording technologies have advanced to produce simultaneous spiking data from massive populations, classical statistical methods often lack the computational efficiency required to handle such data. Machine learning (ML) approaches are known for enabling efficient large scale data analyses; however, they typically require massive training sets with balanced data, along with accurate labels to fit well. Additionally, model assessment and interpretation are often more challenging for ML than for classical statistical methods. 
To address these challenges, we develop an integrated framework, combining statistical modeling and machine learning approaches to identify the coding properties of neurons from large populations. In order to demonstrate this framework, we apply these methods to data from a population of neurons recorded from rat hippocampus to characterize the distribution of spatial receptive fields in this region.
\end{abstract}


\titlepgskip=-15pt

\maketitle

\section{Introduction}
\label{sec:introduction}

The neural encoding problem is a fundamental area of study in systems neuroscience ~\cite{paninski2007statistical}.  
It focuses on understanding the relationship between activity in a neural population and the stimuli or behaviors which influence it. What features of a cognitive process are encoded in an individual neuron or the population of neurons? If multiple features are encoded, what is the relative importance of each? Which neurons are sensitive to which coding variables? These are critical questions that require statistically powerful and computationally efficient methods to address. In general, neural encoding is focused on understanding how information maps from a stimulus or behavioral signal to neural responses and on building models to predict representational spaces \cite{paninski2007statistical, dayan2003theoretical,johnson2000neural, rieke1999spikes}. 
For instance, neural models have been used successfully to relate responses in CA1 region of rat hippocampus during spatial navigation tasks to features of the rat's movement trajectory, the phase of the ongoing theta rhythm in the local field potentials, and the neuron's past spiking history \cite{barbieri2004dynamic, brown1998statistical, zhang1998interpreting, huang2009decoding, o1971hippocampus, frank2002contrasting, truccolo2005point}.

The challenge of neural encoding has increased in recent years as a result of a number of trends in experimental neuroscience. First, there has been a shift from low-dimensional, experimentally controlled stimuli and behaviors to high-dimensional, naturalistic ones, requiring encoding models that can capture the simultaneous, interacting influences of multiple covariates. Second, large-scale electrode arrays and new brain imaging technologies now allow experimentalists to record from hundreds to thousands of neurons or brain sources simultaneously, requiring encoding models that are computationally efficient for high dimensional responses. Third, electrophysiology experiments have moved from a regime where neurons with particular coding properties were specifically targeted, to untargeted preparations where the distribution of cell coding properties can be completely different. This has led to a situation where existing datasets for which coding properties are well studied are unbalanced compared to modern, large-scale datasets \cite{urai2022large, allen2022massive, paninski2018neural, steinmetz2018challenges, vu2018shared}.

The neural encoding problem has been largely addressed statistically \cite{paninski2007statistical}. However, recent challenges make it more difficult to analyze neural data efficiently through a purely statistical framework. One potential approach to resolve this is based on machine learning algorithms. In this paper, we explore existing ML approaches, particularly deep network architectures, and the methods to integrate them with statistical models to address specific modern challenges with neural encoding analyses.

\section{Related Work and Limitations}
\label{sec:introduction}
The problem of neural encoding is fundamentally statistical since neural responses are stochastic \cite{paninski2007statistical}. Statistical methods are used to compare the likelihood of observing particular spike patterns in individual neurons or populations across different neural encoding models with distinct coding properties. These models can include multiple classes of influences, including biological and behavioral signals, the neuron's own past spiking history, and the influences of other neurons in the population \cite{truccolo2005point}. For instance, statistical models for neurons in the CA1 region of rat hippocampus have been used to model spatial place field properties, theta rhythmicity and precession \cite{o1971hippocampus, frank2002contrasting}, and the influence of the neuron's past spiking history. 

Point process models are a class of statistical models that has been successfully used to characterize the factors that influence spiking activity in individual neurons or neural populations. It has been shown that a point process models can be efficiently fit to neural spiking data using a generalized linear model (GLM) framework \cite{truccolo2005point}. Point process GLMs allow researchers to identify significant influences systematically, they provide powerful analysis tools for assessing goodness-of-fit and model refinement, and their parameters are generally interpretable~\cite{paninski2007statistical, truccolo2005point,pawitan2001all,santner1989statistical,zoltowski2018scaling,sarmashghi2021efficient}. However, statistical models are computationally limiting for large scale data analysis due to the fact that the model refinement process is typically performed individually for each neuron. Another common challenge occurs whenever model covariates are able to separate events and non-events perfectly, which leads to infinite maximum likelihood solutions for which computational algorithms can iterate indefinitely \cite{farhoodi2021problem}. Moreover, these models typically make assumptions about the form of the receptive field that must be assessed. If incorrect, these assumptions could lead to bias or increased variability in the model fits and incorrect statistical inference about coding properties. 

The growth of experimental methods in which more neurons are recorded for longer periods of time, necessitates the development of new data analysis methods that are computationally efficient \cite{urai2022large, allen2022massive, paninski2018neural, steinmetz2018challenges, vu2018shared}.
Recently, the advancement of machine learning (ML) algorithms, particularly deep neural networks (DNNs), makes it possible to analyze large scale datasets with high computational efficiency and fewer or no modeling assumptions. These approaches have come to dominate several applications ranging from perceptual, visual object \cite{he2016deep} and auditory speech recognition \cite{sak2014long}, to cognitive tasks, machine translation \cite{luong2014addressing, wu2016google}, motor control tasks such as playing computer games or controlling a robot arm \cite{lange2010deep, mnih2015human}, and so on \cite{kietzmann2019deep}. Deep learning (DL) has recently found its way back into computational neuroscience \cite{kietzmann2019deep} and has been very successful across many applications such as encoding retinal ganglion cells' responses to natural scenes \cite{mcintosh2016deep, batty2016multilayer}, neural encoding and decoding of human visual cortex \cite{wen2018transferring, wen2018neural, horikawa2013neural}, monkey primary motor cortex, somatosensory cortex, and the rat hippocampus \cite{benjamin2018modern, glaser2020machine, tampuu2019efficient}.  
Despite the high predictive power of DL networks, they are parameter-rich models and often need large amounts of data to be adequately trained. Moreover, these models have a black-box nature that leads to difficulty of assessment and interpretation \cite{kietzmann2019deep, batty2016multilayer}. We summarize the pros and cons of statistical methods and ML approaches in Table \ref{table1}. 
The prime use of DL methods in neuroscience is for addressing neural decoding problems \cite{wen2018neural, horikawa2013neural,glaser2020machine, tampuu2019efficient, saeidi2021neural, livezey2021deep, cruz2014neural,chen2021neural, mathis2020deep, iqbal2019decoding} and there is a relative paucity of these approaches devoted to neural encoding and classification \cite{mcintosh2016deep, batty2016multilayer, wen2018transferring, benjamin2018modern, nayebi2022mouse}.

\begin{table}
\caption{Properties of statistical methods and machine learning models}
\label{table1}
\setlength{\tabcolsep}{3pt}
\begin{tabular}{|p{120pt}|p{50pt}|p{50pt}|}
\hline
\textbf{Desired Features}& 
\textbf{Stat Methods}& 
\textbf{ML Models}\\
\hline
Controllable statistical power & 
\checkmark& 
$\times$ \\
\hline
Interpretability& 
\checkmark& 
$\times$ \\
\hline
Model Assessment tools& 
\checkmark& 
$\times$ \\
\hline
Computational efficiency in higher dimensional systems& 
$\times$ & 
\checkmark\\
\hline
Model free & 
$\times$ & 
\checkmark\\
\hline
\end{tabular}
\label{table1}
\end{table}

There are a number of major challenges associated with classifying neural coding properties with ML methods. 1) Both neural spike trains and the signals they encode are time-series data. Deep learning algorithms \cite{lecun2015deep} have been successful in various classification tasks which motivated the recent utilization of DL models for time series classification (TSC) \cite{wang2017time}. However, neuroscience experiments have shifted from completely controlled stimuli, for which simple DL approaches may have worked well, to uncontrolled and naturalistic stimuli, which require DL methods that can capture complex associations between multivariate time series.
2) In order to train deep network classifiers, we need large training datasets along with labels. The data size for training DNNs is often on the order of magnitude of thousands to millions of samples, however the size of data recorded in neuroscientific experiments is often on the order of tens to hundreds of cells. While experimental methods in neuroscience have evolved to include more neurons, there has been a simultaneous increase in the dimensionality and length of the recorded data, making ML classification more difficult.
3) Some classes of neurons or receptive field structures are rarely observed even in large datasets, which leads to imbalance in the training datasets. This is in part related to historical trends in electrophysiological experiments in which electrodes were localized in the brain to target specific cell types and largely excluded other cells that lacked the desired coding properties. For instance, experiments that target place cells in CA1 region of hippocampus, rarely include non-place specific cells. However, recent trends in electrophysiology have led to less targeted experiments in which electrodes with many contacts record from large neural populations with coding properties substantially different from those for which previous robust electrophysiology datasets exist. 
In addition to these structural challenges, previous studies incorporating ML and statistical models have treated these as competing methods, seeking to demonstrate the superiority of one over another in specific applications \cite{chen2021neural, benjamin2018modern, mcintosh2016deep}. Our goal in this work is to demonstrate that classical statistical methods and supervised machine learning algorithms have complimentary strengths and can be used together to address the limitations of each method on its own.

In this study, we address these challenges by exploring an analysis framework that integrates statistical modeling and DL approaches. 1) We implement a CNN-based DL classifier named multi channel deep CNN (MCDCNN) that applies a one-dimensional convolutional channel to each coding feature independently, then combines the learned information across all channels to perform classification. 2) We utilize well-developed statistical methods for neural encoding to generate labels for the training datasets. 3) We use the generative property of statistical models to generate as many samples and labels as needed to increase the training data size. 4) We specifically generate samples to augment data for the minority classes in order to balance the training data. 
We demonstrate the application of this integrated framework on simultaneously recorded local field potential (LFP) and spiking data from the CA1 region of hippocampus of a rat performing a memory-guided spatial navigation task \cite{jadhav2016coordinated, tang2017hippocampal}. We use point process GLMs to classify each of the neurons in a training dataset based on whether their firing is influenced by the rat's location, speed, direction of movement, and the instantaneous phase of its theta rhythm. We generate additional, simulated data based on these model fits in order to augment the training dataset, both increasing its size, and providing balance by increasing the number of samples with less common coding properties. Finally, we train a MCDCNN classifier and evaluate its performance classifying neurons in a separate test dataset as a function of the size and balance of the training data.

\section{Modeling Background}
We begin by reviewing the statistical model identification approach based on point process GLMs. Then we discuss several neural network architectures for the purpose of classification in neural coding. 

\subsection{Point Process-GLM Framework}
A point process model describes the likelihood of any set of localized events occurring in continuous time. Point process models are often used to identify the signals that influence spiking activity in individual neurons or neural populations~\cite{truccolo2005point}. Point processes can be characterized by their conditional intensity function~\cite{daley2003introduction}.
\begin{equation}
    \label{eq:CIF}
    \lambda(t|H(t)) = \lim_{\Delta \to 0}\frac{P[N(t+\Delta)-N(t)=1|H(t)]}{\Delta}
\end{equation}
where $N(t)$ is number of spikes in time interval $(0,t]$ and $H(t)$ is the past spiking history of the neuron or population up to time $t$. For small $\Delta$, $\lambda(t|H(t))\Delta$ is approximately the probability of observing a single spike in the time interval $(t,t+\Delta]$ given the spiking history ~\cite{sarmashghi2021efficient}.

A point process neural coding model defines $\lambda(t|H(t))$ as a function of a set of covariates influencing spiking. A common GLM expresses the log of the conditional intensity of each neuron as linear combinations of functions of extrinsic covariates related to spiking, the neuron’s own spike history, and the past spiking activity of other simultaneously recorded neurons ~\cite{truccolo2005point}. 
\begin{equation}
    \label{eq:CIF-GLM}
    \log \lambda(t|H(t)) = \sum_{i=1}^{p}\theta_i g_i[\nu(t)]
\end{equation}
Where the $g_i(\cdot)$ are a set of basis functions that act on the covariate vector $\nu(t)$, and $p$ is the dimension of the model parameter vector $\theta$. 
GLMs can flexibly capture nonlinear relationships between stochastic signals in a computationally efficient and robust way, and provide powerful tools for assessing goodness-of-fit, and model refinement ~\cite{truccolo2005point,paninski2007statistical,pawitan2001all,santner1989statistical, sarmashghi2021efficient}.

\textbf{Maximum Likelihood Parameter Estimation}
Once an encoding model is expressed as a point process GLM with a log link function as in Eq.\eqref{eq:CIF-GLM}, the likelihood surface is guaranteed to be convex, ensuring that there exists a unique maximum \cite{truccolo2005point, sarmashghi2021efficient}. The maximum likelihood estimator for the model parameters can be computed using an iteratively re-weighted least squares (IRLS) procedure ~\cite{truccolo2005point}. The IRLS procedure also computes the observed Fisher information matrix, $I_{\hat{\theta}}$, which allows for the construction of confidence intervals and standard tests of significance of the influence of individual or multiple covariates~\cite{truccolo2005point, santner2012statistical, pawitan2001all, sarmashghi2021efficient}.

\subsection{Deep Learning Time Series Classification}
Neuronal spiking activity and the signals that influence it are time series data. There are multiple neural network architectures that are appropriate for time series classification (TSC) tasks. In this section, we provide some background materials regarding TSC classification and explore various neural network architectures for this task. We start with a simple multi layer perceptron (MLP) and then discuss more complicated architectures such as convolutional neural networks (CNNs) and recurrent neural networks (RNNs).

\subsubsection{Time Series Classification}
\begin{figure}[!t]
\centerline{\includegraphics[width=\columnwidth]{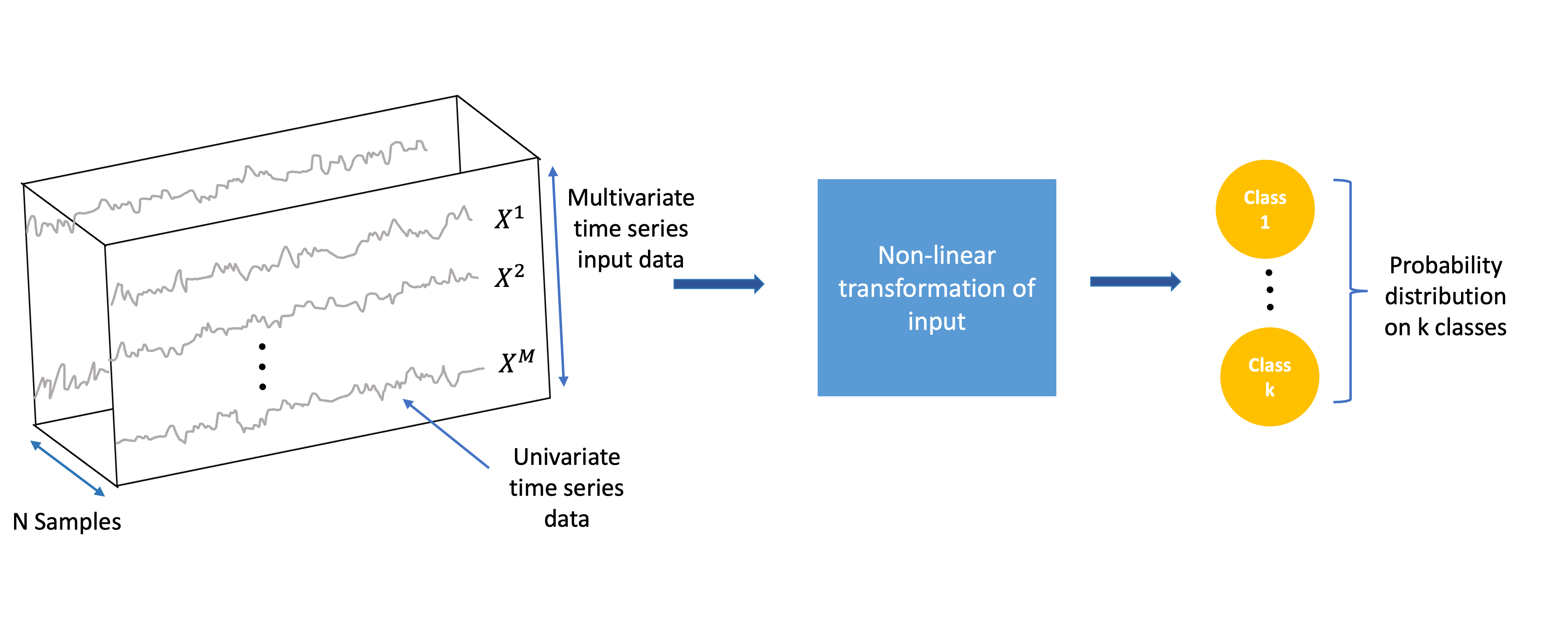}}
\caption{General deep learning framework for multivariate time series classification}
\label{fig1}
\end{figure}

Time series classification is among the most challenging problems in data mining \cite{yang200610}. Deep learning algorithms \cite{lecun2015deep} have been successful in various classification tasks which motivated the recent utilization of DL models for time series classification (TSC) \cite{wang2017time, fawaz2019deep}. A general deep learning framework for a multivariate time series classification (MTSC) is illustrated in Fig. \ref{fig1}. A multidimensional time series, $X = [X^1, X^2, ..., X^M]$ consists of $M$ univariate time series $X^i = [x^i_1, x^i_2, ..., x^i_T]$ where $T$ is the number of time steps. 
Let $D = \{(X_1,Y_1), (X_2,Y_2), ..., (X_N,Y_N)\}$ be a dataset where $Y_i$ is a one-hot label vector, and $X$ and $Y$ represent the input and output spaces respectively. 
The main goal is to train a classifier using a dataset D to map from the input space $X$ to a probability distribution of the labels \cite{wang2017time, fawaz2019deep}. In neural coding, the input variable consists of the activity of a neuron along with the biological and behavioral signals that influence it, and the output variable provides a label that defines the coding properties for that neuron. Classification depends not on the statistical properties of the neural activity or covariates on their own, but on the dependence between these sets of variables across time. 

\subsubsection{Multi Layer Perceptrons}
MLPs or fully-connected (FC) feed-forward networks are the simplest and most traditional deep learning network, which are considered as the baseline architecture for time series classification \cite{wang2017time, fawaz2019deep}. A MLP has three major layers of neurons: the input layer, hidden layer, and output layer, and each neuron in one layer is connected with a specific weight to every neuron in the following layer through an activation function. In this feedforward network architecture, learning can be carried out through the backpropagation algorithm \cite{friedman2017elements, rosenblatt1961principles, rumelhart1985learning}. Despite the fact that MLPs do not exhibit temporal invariance meaning each time step has its own weight and the temporal information is lost \cite{fawaz2019deep}, it still provides a useful baseline for the TSC task. 

\subsubsection{Convolutional Neural Networks}
Convolutional neural networks (CNNs), particularly deep CNNs, have been very successful in many domains of classification \cite{lecun2015deep, krizhevsky2012imagenet, szegedy2015going}. The successes of CNN architectures motivated researchers to adopt them for time series analysis as well \cite{gamboa2017deep}. Convolutional networks are practically feedforward neural networks that use convolution instead of general matrix multiplication. CNNs work with fixed-size, spatially-organized data. The features from the input space can be extracted by a set of convolution layers applying weight-sharing filters and dimension-reducing pooling. Unlike images, where two-dimensional filters (width and height) are used, time series classification can be implemented using a one-dimensional CNN with the filters sliding over time \cite{weytjens2020process, fawaz2019deep}. Next, the outputs of the feature extraction are fed to a fully connected network to estimate a probability distribution over the class variables. CNNs achieve their high levels of success by leveraging three ideas: 1) sparse weights due to the use of kernels (filters) smaller than the input size, which leads to computational efficiency; 2) sharing weights across spatial or temporal locations, leading to less memory usage and less computational effort; 3) translation invariance, which allows for robust classification of similar, temporally shifted input patterns  \cite{Goodfellow-et-al-2016}. 

\subsubsection{Multi Channel Deep CNNs}
\begin{figure}[!t]
\centerline{\includegraphics[width=\columnwidth]{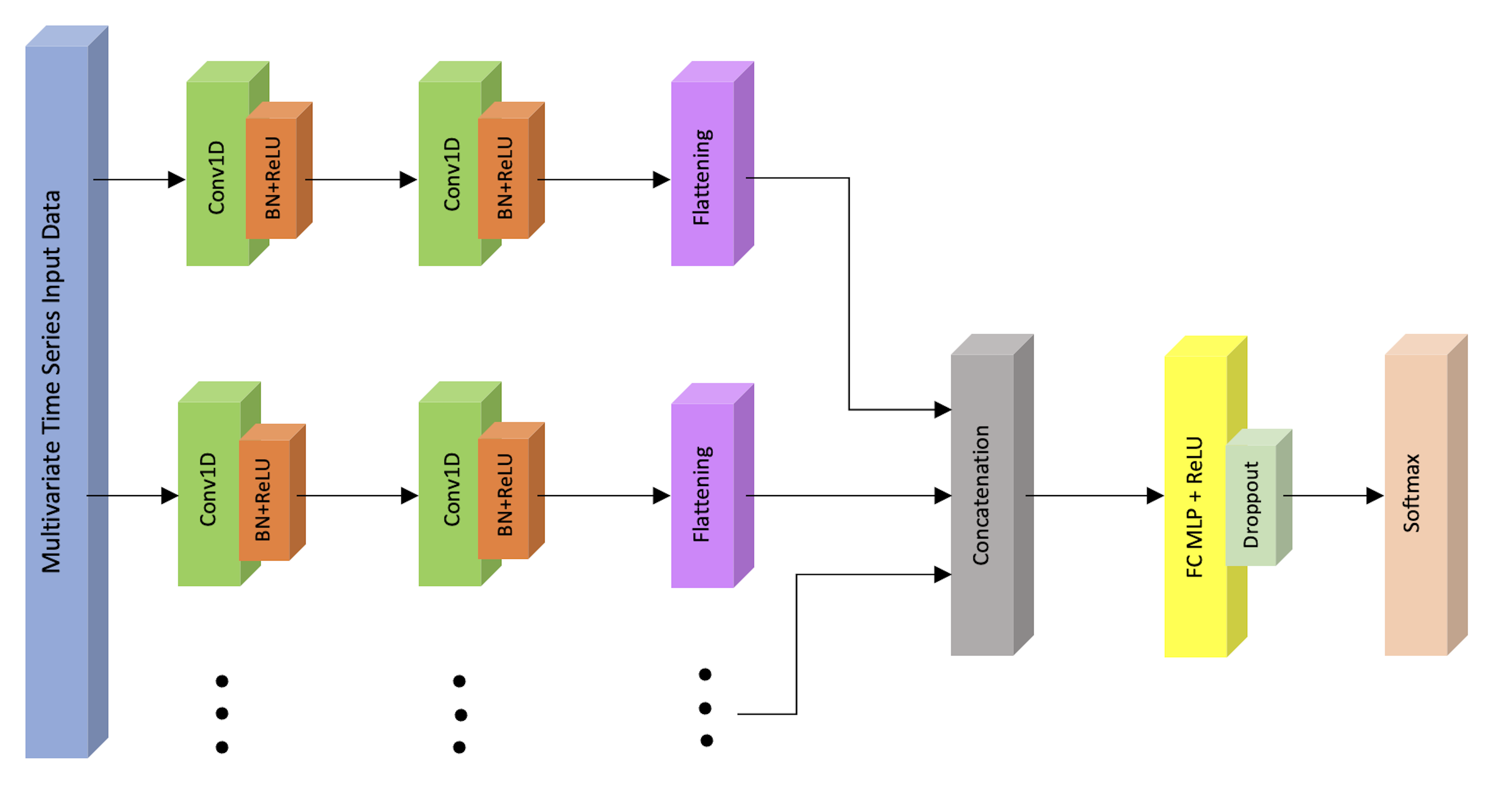}}
\caption{Multi channel deep convolutional neural network (MCDCNN) architecture}
\label{fig2}
\end{figure}

Multi channel deep convolutional neural networks (MCDCNNs) are modified CNNs that are well-suited for MTS data types. These models first learn features from each input dimension in each channel, then combine information across all channels and feed them to a fully connected network for classification. In other words, the convolutions are applied independently on each dimension of the input data to learn the features (Fig. \ref{fig2}). These networks were originally proposed for two variable time series data \cite{zheng2014time, zheng2016exploiting, fawaz2019deep}. 

\subsubsection{Recurrent Neural Networks}
Recurrent neural networks (RNNs) are a family of neural networks that are designed for processing sequential data \cite{rumelhart1986learning}. They are able to handle temporal structure in time series data, and they are specialized to handle sequences of variable length \cite{weytjens2020process}. A RNN processes a sequence of events, one time step at a time. For each time step, it passes information about the current and previous time to the next time step, until reaching the last one whose output is propagated to the next layer \cite{weytjens2020process}. RNNs are particularly designed to predict an output for each element (time step) in the time series \cite{langkvist2014review} and are thus often used for time series forecasting, but are less often used for time series classification \cite{fawaz2019deep}. They typically suffer from the vanishing/exploding gradient problem when trained on long time series \cite{bengio1993problem, bengio1994learning, pascanu2013difficulty}. RNNs are also considered hard to train and parallelize  \cite{pascanu2013difficulty}. However, these networks' ability to capture temporal structure in time series data make them worth considering for neural encoding. The problem with vanishing/exploding gradients can be addressed by a particular form of RNNs called long-short term memory (LSTM) \cite{hochreiter1997long}. The LSTM model uses self-loops to produce paths where information related to the gradient can flow for long stretches of time \cite{Goodfellow-et-al-2016}. LSTMs have been successful in the neural coding domain particularly for predicting neural responses to stimuli \cite{mcintosh2016deep, batty2016multilayer} and for neural decoding \cite{glaser2020machine, tampuu2019efficient}. 

\section{Materials and Methods}

\subsection{Experimental Data} The data has been described in detail previously ~\cite{jadhav2016coordinated,tang2017hippocampal}, but we summarize the key points here. One male Long–Evans rat weighing 450–550 g was implanted with a movable array of recording tetrodes in the CA1 region of hippocampus, which recorded simultaneous local field potentials (LFP) and neural spiking activity from 254 neurons. The animal had been trained for 8 days in 2-4, 20-minute run sessions per day to perform a spatial alternation task on a W-shaped track, alternating between the left and right arms before returning to the center arm. It received a reward at the ends of the left and right arms after each correct alternation ~\cite{sarmashghi2021efficient, jadhav2016coordinated,tang2017hippocampal}. This data was recorded at a 1500 Hz sampling frequency and later re-sampled at 100 Hz (10 ms time resolution) prior to data analysis. This data is publicly available at the CRCNS data sharing site.

\subsection{Point Process-GLMs} In this task, the candidate predictor signals are the rat’s linearized- position $x(t)$, speed $s(t)$, direction $d(t)$, and the phase of the theta rhythm $\phi(t)$ (Fig. \ref{fig4}). $x(t)$ represents the distance from the tip of the central arm (in cm), with the range [0,80] representing the center arm, [80,190] representing the right arm, and [190,300] representing the left arm. $d(t)$ is an indicator function that is equal to 0 for outbound movements away from the center arm and equal to 1 for inbound movements toward the center arm. We constructed point process models such that the conditional intensity had the following form:

\begin{equation}
    \label{eq:HippStatModel}
    \log \lambda(t) = \sum_{i=1}^{p}\theta_i g_i(x(t),s(t),d(t),\phi(t))
\end{equation}

where $p$ is the dimension of model parameter $\theta_i$, and $g_i(.)$ represents a set of basis functions acting on the candidate covariates. We start with a null model (class 0) comprising a simple homogeneous Poisson model and add predictors incrementally. The first such model merely includes the linear position signal and uses modified cardinal spline basis functions \cite{sarmashghi2021efficient} to estimate if the neuron has a statistically detectable place field (class 1). This model can be written as $\log \lambda(t) = \theta_0 + \sum_{i=1}^{p_S}\theta_i g_i(x(t))$, where $g_i$ are the cardinal spline basis functions and $p_S$ is the number of spline control points. The value of $p_S$ and the locations of the spline control points were selected by first constructing an occupancy normalized histogram of the place field of each neuron over the linearized track
as in Fig. \ref{fig4}B, and then placing control points on the local peaks on the histogram. The next model class includes an interaction between place coding and the rat's speed at each time step (class 2). This model has the form $\log \lambda(t) = \theta_0 + \sum_{i=1}^{p_S}\theta_i g_i(x(t))I_{s(t)>2}$, where $I_{s(t)>2}$ is an indicator function that is equal to $1$ if the rat's speed is above 2 cm/sec at time $t$ and $0$ otherwise. Next, we added the rat's direction of motion (outbound vs inbound) to the model along with the speed and place coding (class 3). This model has the form $\log \lambda(t) = \theta_0 + \sum_{i=1}^{p_S}\sum_{j=1}^{2}\theta_{i,j} g_i(x(t))I_{s(t)>2}I_{D_j}$, where $I_{D_1}$ and $I_{D_2}$ are indicator functions for inbound and outbound trajectories, respectively. Finally, we added an interaction of the phase of the rat's theta rhythm to the direction, speed and place coding model (class 4). This model has the form $\log \lambda(t) = \theta_0 + \sum_{i=1}^{p_S}\sum_{j=1}^{2}\sum_{k=1}^{4}\theta_{i,j,k} g_i(x(t))I_{s(t)>2}I_{D_j}I_{\Phi_k}$, where $I_{\Phi_k}$ is an indicator function for the phase of the theta rhythm being in each of the following intervals, respectively: $\Phi_1 = [-\pi, -\pi/2)$, $\Phi_2 = [-\pi/2, 0)$, $\Phi_3 = [0, \pi/2)$, and $\Phi_4 = [\pi/2, \pi)$. 

We estimated model parameters using maximum likelihood.

We classify each neuron based on a step-up model identification procedure: for each neuron we begin with the null model and perform sequential maximum likelihood ratio tests to determine if the next set of covariates provide a significant improvement to model fit. The neuron's classification is determined by the first test in this sequence that fails to reach significance. 

\subsection{Simulation Analysis}
Before applying deep learning models to the data, we assessed the classification accuracy of several network architectures on simulated data. We simulated each neuron so as to fall into one of the GLM classes described above. Simulated neurons in model class 0 were generated as a homogeneous Poisson process with constant rate, $\lambda_0(t) = \alpha$, where $\alpha$ for each cell is drawn independently from a uniform distribution with range [0, 200] spike/s. This range is based on minimum and maximum of observed intensity from experimental data. Neurons in class 1 were simulated with an intensity function with a Gaussian shape as a function of the rat's position, $\lambda_1(t) = \alpha * exp((x_t - a)^2/2b^2)$ where $\alpha$ is the peak firing rate at the center of the place field, $a$ is the location of the peak, and $b$ defines the width of the place field. $\alpha$, $a$, and $b$ were drawn from independent uniform distributions with ranges [6, 9] spikes/s, [10, 290] cm (linear position varies between 0-300 cm), and [10, 100] cm respectively. Neurons in class 2 were simulated with a place field that is modulated by the rat's speed, $\lambda_2(t) = \lambda_1(t) * exp(\gamma s_t)$, where $\lambda_1(t)$ is the place field structure from class 1 neurons, described above, $s_t$ is rat's speed and $\gamma$ is a fixed constant set to 0.05. Finally, neurons in class 3 have place fields modulated by speed and direction: $\lambda_3(t) = \lambda_2(t) * exp(\eta dt)$, where $d_t$ is an indicator function that is equal to 1 for inbound movements and -1 for outbound movements and $\eta$ is a fixed parameter set to 0.5 for this analysis.

\subsection{Deep Learning Models}
We explored a number of network architectures to assess their classification performance on our simulated dataset, then we used the results to select one neural network for classifying the real data.  
The input space includes multivariate time series: the spike train, the rats’ movement trajectory including position, direction, and speed, and the instantaneous phase of the theta rhythm in the rat's LFP. The output labels are either binary variables indicating whether a neuron is or is not significantly influenced by a set of covariates, or are 5-ary, indicating which of the model classes discussed in the previous section was selected by the model identification procedure. 

For both the simulation and real data analyses, $70\%$ of the data is devoted to training, $10\%$ to validation, and $20\%$ to testing. We used mini-batch gradient descent (GD) \cite{dekel2011optimal} as the learning algorithm to update model parameters. Mini-batch GD is based on gradient descent (GD); however, instead of updating parameters once per epoch (iteration), it updates them in real time within each epoch. 
This algorithm splits the data to $m$ batches of samples and updates parameters in every batch per epoch, which reduces the variance in the estimate of the gradient and leads to faster converge \cite{khirirat2017mini}. For this analysis we split the training data into 128 batches and iterated the learning algorithm on the entire training set 90 times. All models were trained using RMSprop optimizer which is an extension of stochastic gradient descent (SGD) algorithm  \cite{ruder2016overview} to minimize a categorical cross-entropy loss function \cite{murphy2012machine}. Classification performance was assessed by the classification accuracy, which is the ratio of the number of correct predictions to the number of input samples. We summarize the software configuration, splitting procedure, and learning parameters in Table \ref{table2}. 

\begin{table}
\caption{Software and model configuration and parameters }
\label{table2}
\setlength{\tabcolsep}{3pt}
\begin{tabular}{|p{60pt}|p{60pt}|p{100pt}|}
\hline
\textbf{Software}& 
\textbf{Data Split}& 
\textbf{Training Process}\\
\hline
Python3/3.8.10& 
Training: $70\%$& 
Batch Size: 128\\

TensorFlow/2.5.0& 
Validation: $10\%$& 
Epochs: 90 \\

& 
Testing: $20\%$& 
Loss Function: Cross Entropy\\

& 
& 
Optimizer: RMSprop\\

& 
& 
Performance Metric: Classification Accuracy\\

\hline
\end{tabular}
\label{table2}
\end{table}

Specific implementation details for each of the network architectures explored are provided below.

\subsubsection{Multi Layer Perceptrons (MLPs)}
The MLP network includes five fully connected layers in total, with three hidden layers of 10, 7, and 4 neurons per layer respectively. In the first layer, the product of the input data with a weight matrix is passed through a Rectified Linear Unit (ReLU), $R(z) = max \{0, z \}$, and this procedure is repeated at each hidden layer. The final layer takes weighted inputs from the final hidden layer to determine the classification result using either a sigmoid activation function, $S(z) = 1/(1 + exp(-z))$ , for the binary classification or a softmax, $F(z)_i =  exp(z_i)/\sum_j exp(z_j)$, for categorical classification. 

\subsubsection{Convolutional Neural Networks (CNNs)}
The CNN network consists of five layers in total including three layers of one-dimensional convolutions. The first convolution layer has 10 filters with kernel size 3, and uses a ReLU activation function. The second layer contains 7 filters of kernel size 3 along with a ReLU. Next, MaxPooling is applied to downsample the output of second. The third layer includes 4 filters with size 3 and a ReLU function. Then, GlobalAveragePooling \cite{zhou2016learning} is used to downsample the features extracted from the convolutional layers. The final stage is a FC network with a sigmoid or softmax function applied for the classification output.

\subsubsection{Multi Channel Deep CNNs (MCDCNNs)}
In this network, two layers of one-dimensional convolutions including 4 filters with a kernel size of 3, are applied to each input dimension independently for the real data analysis. For the simulation analysis, we used 3 filters of size 2. The output of the top convolutional layer is passed through a batch normalization operation to help the network converge quickly \cite{ioffe2015batch}, and is then put through ReLU activation functions. The outputs of this process are concatenated across all input dimensions and fed into a FC network with 128 nodes and a ReLU activation. At the end, a dropout layer was used with rate 0.5 to prevent overfitting \cite{srivastava2014dropout, fawaz2019deep}, and a sigmoid or softmax function was used for classification (Fig. \ref{fig2}).

\subsubsection{Long-Short Term Memory (LSTMs)}
The LSTM network consists of three layers in total including two LSTM layers. The first layer has 6 LSTM units which use sigmoid activation functions. The second layer includes 4 LSTM units and a sigmoid function. At the last layer, sigmoid or softmax functions were used for classification.

\subsection{Integration and Data Augmentation}
The neural dataset does not include labels. Putative labels were generated using the statistical model identification procedure described above. These statistical models were also used to augment the training datasets to include more data and to provide balance between classification categories and generate data from rare or unobserved receptive field structures.
When augmenting data to provide balance, we generated spike trains based on methods described in simulation analysis, from neurons with receptive fields according to the models in Eq. \eqref{eq:HippStatModel} using a distribution of parameter values, irrespective of whether similar parameters values are observed in the actual data. We refer to this approach for augmentation as \textbf{balanced augmentation}.

We also generate spike trains from simulated neurons according to Eq. \eqref{eq:HippStatModel} using the estimated parameter values from observed neurons to increase the size of our training sets, without changing the distribution of observed spiking properties. We refer to this augmentation approach as \textbf{empirical augmentation}.

\section{Results}
\begin{figure}[!t]
\centerline{\includegraphics[width=\columnwidth]{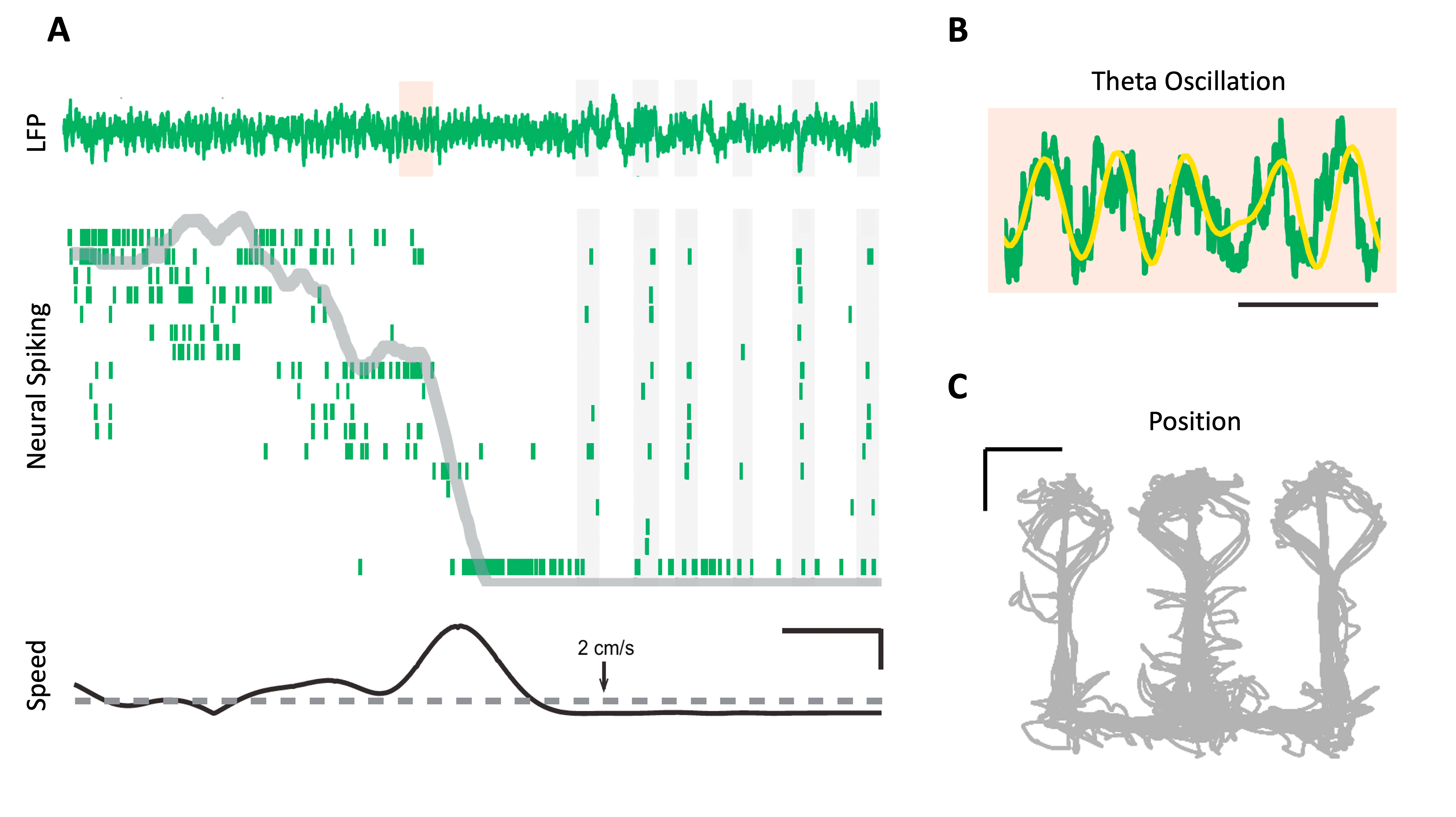}}
\caption{\textbf{Hippocampal neural activity patterns during behavior}. (A) From top to bottom, LFP (1-400 Hz) activity in CA1 and spiking activity of 18 CA1 neurons represented in a raster plot in green. The gray line indicates the linearized position of the animal. The bottom plot shows the rat's speed vs a threshold speed of 2 cm/s, used to differentiate the running state from rest. The scale bars are 2s (horizontal) and 10 cm/s (vertical). (B) Magnification of the shaded area from A where the yellow line is filtered LFP in the theta rhythm bandwidth (6-12 Hz). The scale bar is 250 ms. (C) The animal's movement trajectory in a W-track maze. The scale bar is 20 cm. (Jadhav et al. 2016)}
\label{fig3}
\end{figure}

To demonstrate the interplay between statistical models, simulation, and ML, we have applied an integrated approach to the problem of classifying neurons from the CA1 region of rat hippocampus during a spatial memory and navigation task based on their coding properties. It is well known that rat hippocampus contains place cells, which code for the rats’ location during spatial navigation, but hippocampal neurons have also been shown to have activity that depends on movement speed and direction, the phase of the theta rhythm, and on the timing of past spikes \cite{o1971hippocampus, frank2002contrasting, truccolo2005point}. 
Fig. \ref{fig3} shows an example of the data for a small subset of neurons over a short movement interval. Fig. \ref{fig3}A shows the LFP (1-400 Hz) and the spiking activity of 18 CA1 neurons in a raster plot. The linearized location of the animal is shown as a gray line in the raster plot. The bottom panel shows the animal's speed along the W-track and a 2 cm/s threshold to differentiate the running state from the rest state. Fig. \ref{fig3}B shows a magnification of the filtered LFP signal in theta range (6-12 Hz) in the shaded area from from the top panel of Fig. \ref{fig3}A. The rat's movement trajectory through the W-track maze is shown in Fig. \ref{fig3}C. 

\begin{figure}[!t]
\centering{\includegraphics[width=\columnwidth]{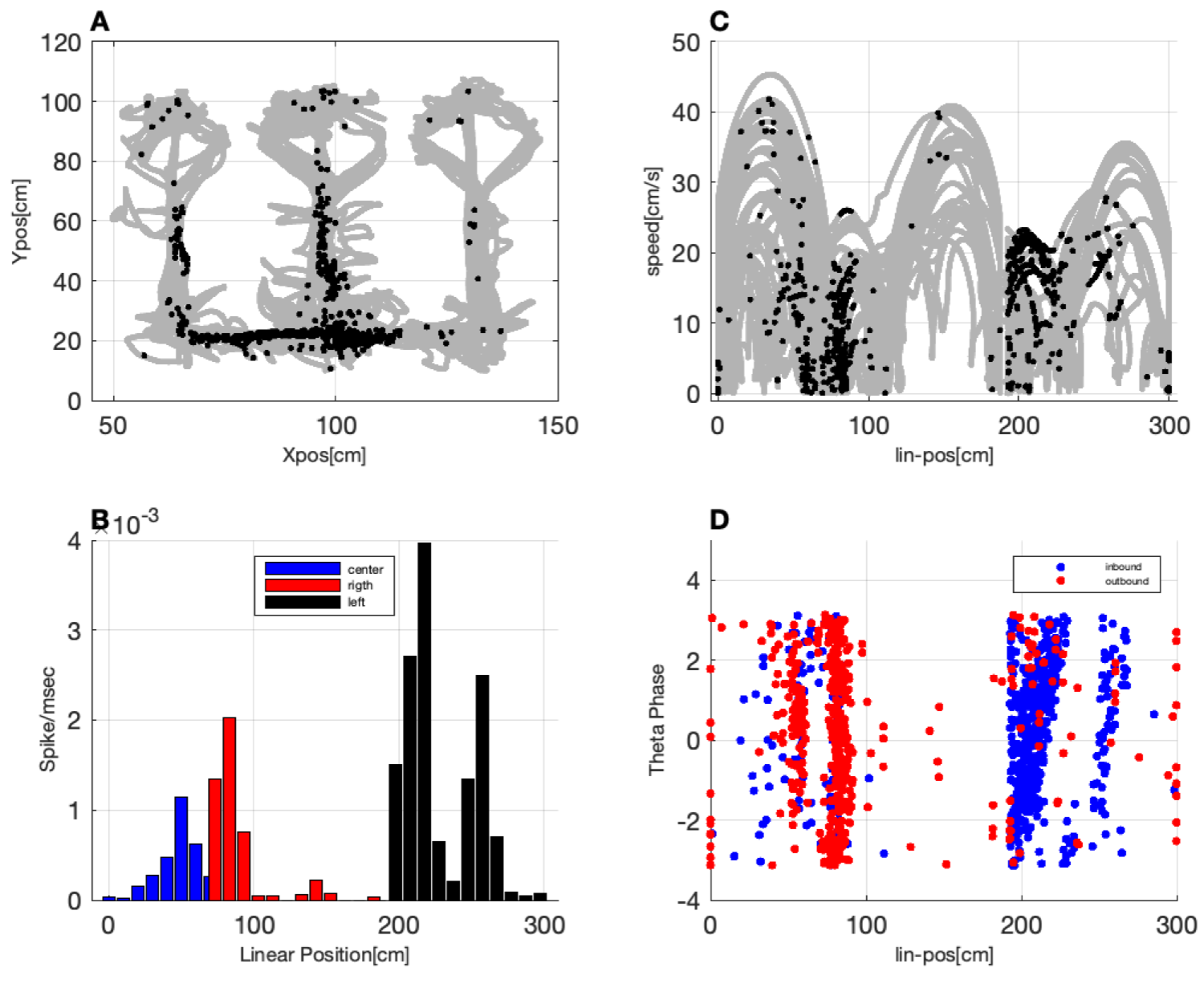}}
\caption{\textbf{Coding properties of an individual neuron in CA1}. (A) Locations of rat at spike times are shown as black dots within the rat's movement trajectory in gray. (B) Occupancy-normalized histogram
of the neuron's firing in a linearized version of the track. Each color
represents a different arm of the W-track. (C) Position and speed of the rat during movement (gray) and at spike times (black). (D) Position and theta phase during spike times during inbound (blue) and outbound (red) trajectories.}
\label{fig4}
\end{figure}

A visualization of some of the coding properties of an example hippocampal neuron is shown in Fig. \ref{fig4}. Fig. \ref{fig4}A shows the rat's movement through the W-track in gray and the location of the animal when the neuron fired as black dots. This neuron tends to be more active on the left side of the track. Fig \ref{fig4}B shows an occupancy normalized histogram of the place field of this neuron on a linearized version of the track, where each color represents a different arm. As expected based on \ref{fig4}A, the highest firing rate occurs on the left arm of the track, but there is also a smaller place field at the beginning of the right arm. Fig. \ref{fig4}C suggests an interesting interaction between place and speed coding in this neuron. Again, the rat's movement trajectory is shown in gray and the location and speed of the rat when the neuron spikes is shown in black. When the rat is in this neuron's primary place field on the left arm of the track (around 200 cm in the linearized position), the neuron fires more when the rat is running quickly (10-20 cm/s). When the rat is moving toward the right arm of the track around 80 cm in the linearized position), this neuron also spikes, but generally when the rat is moving more slowly (0-10 cm/s).  Fig. \ref{fig4}D shows an interaction between the rat's position and the phase of its theta rhythm in determining neural spiking. The blue dots represent the rat's position and theta phase during spiking when the rat moves inbound from the outer arms to the center and the red dots correspond to the rat's position and theta phase at spike times when the rat moves outbound from the center to the outer arms. We see that the primary place field on the left arm occurs during inbound trajectories while the smaller place field in the right arm occurs during outbound trajectories. We also note that the blue dots are clearly slanted, with spikes occurring at earlier phases of the theta rhythm as the rat moves further into the neuron's place field. This phenomenon is known as theta precession \cite{o1993phase, skaggs1996theta}.

\begin{figure}[!t]
\centering{\includegraphics[width=\columnwidth]{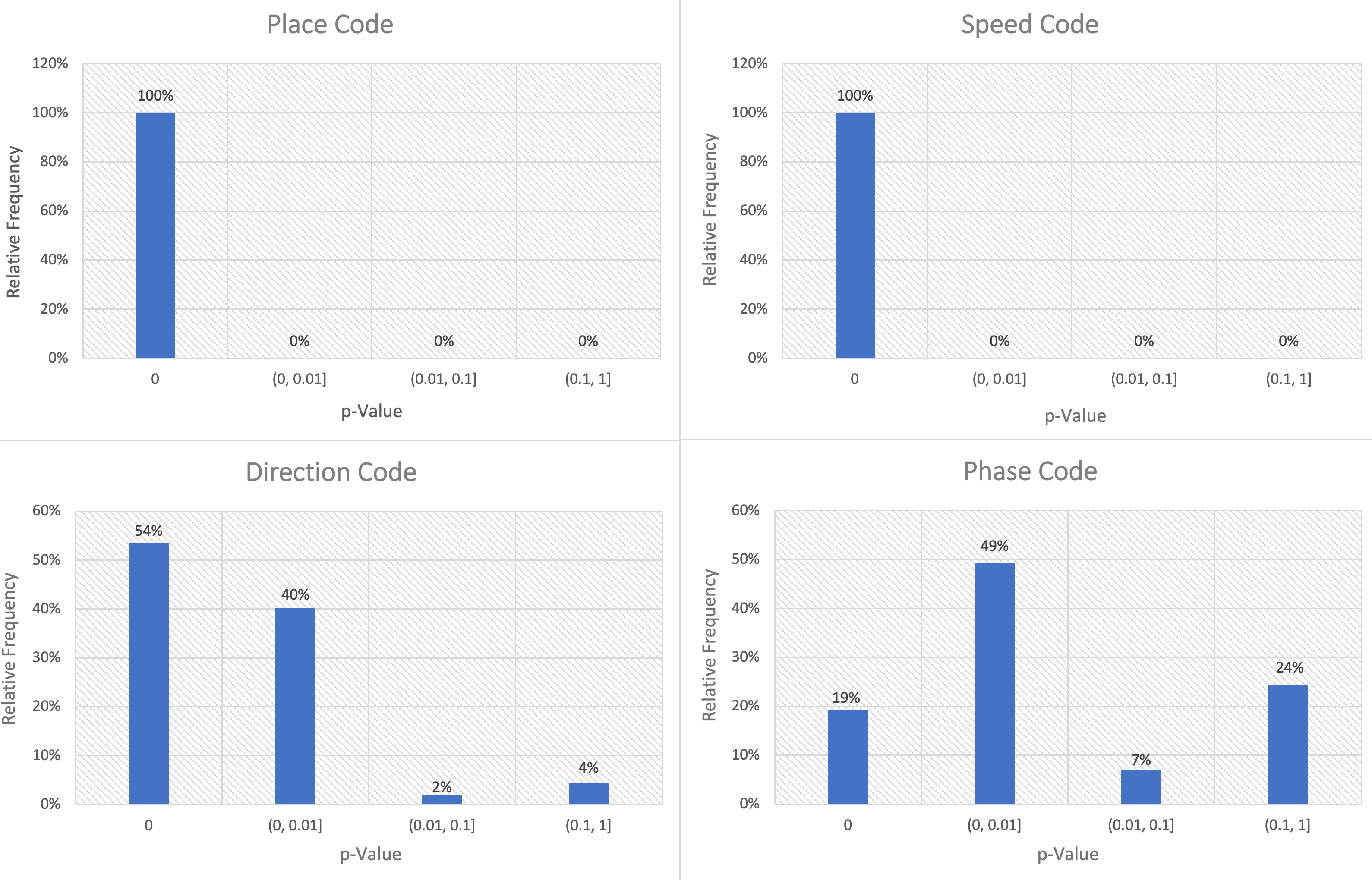}}
\caption{\textbf{Population summary of coding properties
in CA1 population}. Relative frequency of p-values for maximum likelihood ratio tests for each potential predictor of spiking, including position, speed, direction and theta phase.} 
\label{fig5}
\end{figure}

We applied a statistical model identification framework to classify the coding properties of a population of 254 hippocampal neurons. As explained above, we start with a null model in which neurons fire as a homogeneous Poisson process, and add potential predictors of neural spiking to the model sequentially, starting with position, followed by direction, speed, and theta phase. Each model class is selected over the previous one if adding the new predictors leads to a p-value below 0.01 using a maximum likelihood ratio test. Testing occurs sequentially, so that cells that are not found to have significant position dependence are not evaluated for any other influences, for example. Fig. \ref{fig5} demonstrates the population level summary statistic of the candidate variables versus p-values computed from the maximum likelihood ratio tests. In this dataset, every single cell has significant place coding, which is not surprising considering the role rat hippocampus plays in spatial navigation, and the fact that in this experiment, tetrodes were implanted specifically to find place cells. Slightly more surprising is the fact that every neuron's spiking is also influenced significantly by the rat's speed, even when accounting for dependence between position and speed in this task. Using this model identification paradigm, we also found that $94\%$ of the neurons code for direction, and $68\%$ of the population code for theta phase. We used these classification results to label the data to train our supervised machine learning algorithms. 

\begin{figure}[!t]
\centering{\includegraphics[width=\columnwidth]{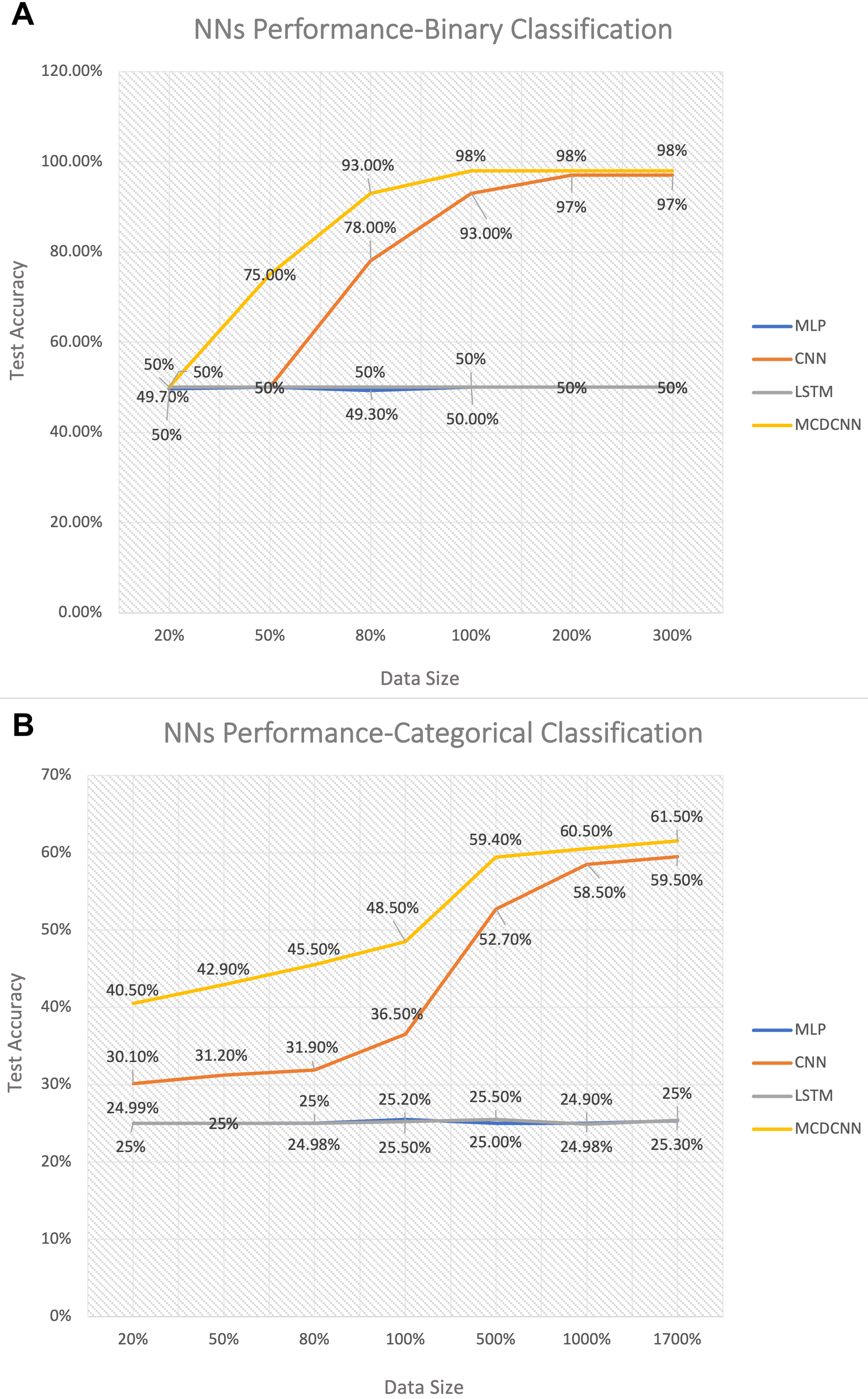}}
\caption{\textbf{Performance of different neural network architectures on classifying simulated data}. Performance of four neural networks including MLP, LSTM, CNN, and MCDCNN for (A) binary classification of place field vs non-place specific firing where $100\%$ data size means 204 neurons, and (B) categorical classification with 4 classes including place, speed, and direction specific firing. $100\%$ data size means 816 neurons.}
\label{fig6}
\end{figure}

Before applying the ML classifiers to this data, we begin with a simulation study to assess the strengths and weaknesses of our proposed architectures. For this analysis, we begin with a simple binary classification in which the inputs are spike trains and the rat's linearized movement trajectory, and output is whether each neuron codes for position (class 1) or not (class 0). We vary the data size from 20\% the size of our dataset (50 neurons) to 300\% its size (750 neurons). For each dataset, we divide the data into a training set using 80\% of the neurons, where half of those neurons are place specific and the other half are non-place specific, and a test set of the remaining 20\% of the neurons. Fig. \ref{fig6}A shows the classification accuracy of each of the network architectures for various training set data sizes. $100\%$ data size means that the training set includes 204 neurons and any smaller percentage indicates that the training set was reduced to that fraction of its original size, maintaining balance between both classes. We find that for this training set size, only the CNN and MCDCNN architectures are able to classify the test set at all. Then, we doubled ($200\%$) and tripled ($300\%$) the training size while keeping the balance between classes in order to illustrate the classification accuracy of MCDCNN and CNN classifiers as a function of training set size. 

As Fig. \ref{fig6}A shows, the MCDCNN's test accuracy grows above chance at a smaller data size compared to the CNN. Once the training set size is doubled to around 400 neurons, the CNN is becoming asymptotically equivalent to the MCDCNN curve, with both architectures providing high accuracy classification. We had expected the LSTM architecture to be a strong candidate to provide efficient classification due to presence of temporal structure in data. However, this network was not able to classify above change levels for the data sizes we explored. We also expected the MLP to work well for a simple binary classification, which also proved incorrect. These result suggest that even though these architectures have been used previously to classify multivariate time-series datasets, they may not be appropriate for problems that involve identifying dependence structure between multiple time series.

Next, we expanded the simulation analysis to a categorical classification problem where the input variables included the spike trains, rat's position, speed, and direction. The outputs are indicators for each of the four model classes: no coding properties (class 0); place code only (class 1); place and speed code only (class 2); and place, speed, and direction code (class 3). We used the same approach as the binary case by generating balanced simulated data of different sizes, in this case ranging from 160 neurons to 13,600 neurons. In each training set, there were an equal number of simulated neurons in each model class. Fig. \ref{fig6} (B) illustrates test accuracy of different neural networks as a function of training size where $100\%$ data size means 816 neurons (204 per class). Again, the MLP and LSTM were unable to classify, and the MCDCNN and CNN grow in classification accuracy as the training size increases. Again, the MCDCNN performs better an any fixed training set size compared to the CNN, but the CNN becomes asymptotically equivalent to the MCDCNN with sufficiently large datasets. Based on these results, we opted to focus on the MCDCNN architechture in our analysis of the real neural data.

\begin{figure}[!t]
\centering{\includegraphics[width=\columnwidth]{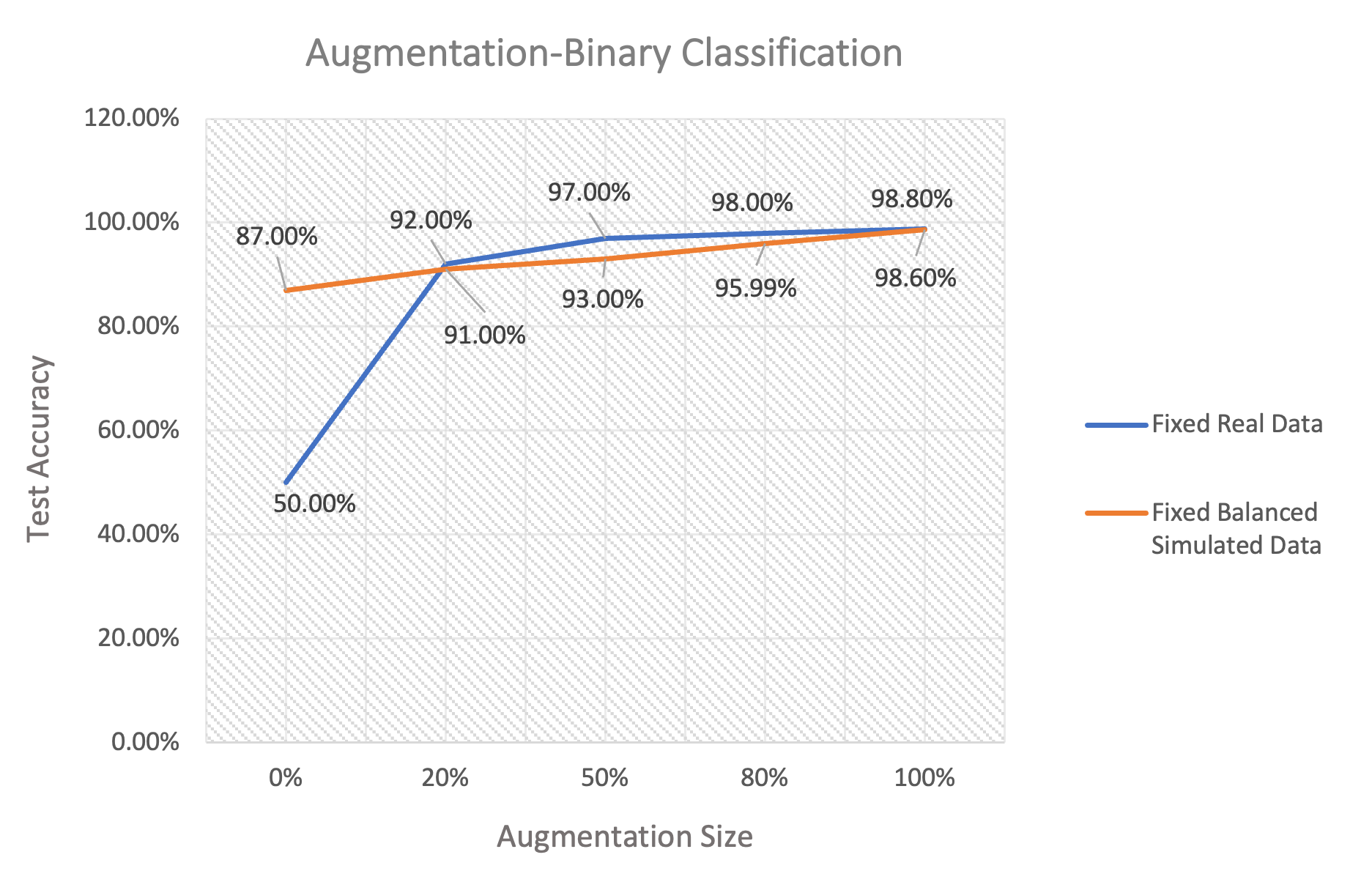}}
\caption{\textbf{Test accuracy of binary classification for two training scenarios}. Scenario (1): fixed amount of real data that includes only place specific neurons (class 1) augmented by adding balanced data of size $20\%$, $50\%$, $80\%$ and $100\%$ of the original data size, 204 neurons, (blue line). Case (2): fixed amount of balanced simulated data augmented by adding subsets of the real dataset of different sizes (orange line). }
\label{fig7}
\end{figure}

We begin the data analysis with a binary classification example corresponding to the simulation analysis procedure, which included only class 0 and 1. Since the real data contains no examples of non-place specific activity (class 0), we augmented it with balanced simulated data. Then, we divided the data to training and test sets where the test set had a fixed size of 50 neurons from the real data (class 1) and 50 neurons from balanced simulated data (class 0). For the training set, we considered two augmentation scenarios. In the first, the amount of real data in the training set was fixed at 204 neurons, and we added different amounts of balanced simulated data (blue line in Fig. \ref{fig7}). The x-axis in this figure indicates the size of the balanced simulated data added to the training set as a fraction of the original training data size. Thus a value of $100\%$ indicates that 204 balanced simulated neurons were added to the 204 real neurons to generate the training set. The second augmentation scenario consists of a fixed amount of simulated data that includes 204 neurons, 102 neurons in class 0 and 102 neurons in class 1, to which we add data from varying numbers of real neurons (orange line in Fig. \ref{fig7}).

When there is no augmentation to the real data (blue line, 0\% augmentation), the classification accuracy on the test set is 50\%, since the data is imbalanced and the model in only trained on one class. When we add $20\%$ of the simulated data (class 0) to balance the training set, the classification accuracy increases to $92\%$, suggesting that a relatively small amount of balanced simulation data can drastically improve the value of an unbalanced training set. Adding mode balanced simulation data then leads to modest additional improvements, which asymptote towards $100\%$ classification accuracy. 
The second simulation scenario shows how the results would change if the training dataset were balanced from the outset (orange line, 0\% augmentation). Rather than starting at chance, the classification accuracy for a balanced dataset of the same size as the real data is $87\%$. We then add the unbalanced real data to this initially balanced training set and the classification accuracy quickly converges with the curve from simulation scenario 1. Taken together, the results from these two scenarios suggest a couple of things. First, that even a small amount of balanced data can substantially improve the quality of in imbalanced training set. Second, that once a training dataset is relatively balanced, adding unbalanced data leads to about the same improvement that adding balanced data does.

\begin{figure}[!t]
\centering{\includegraphics[width=\columnwidth]{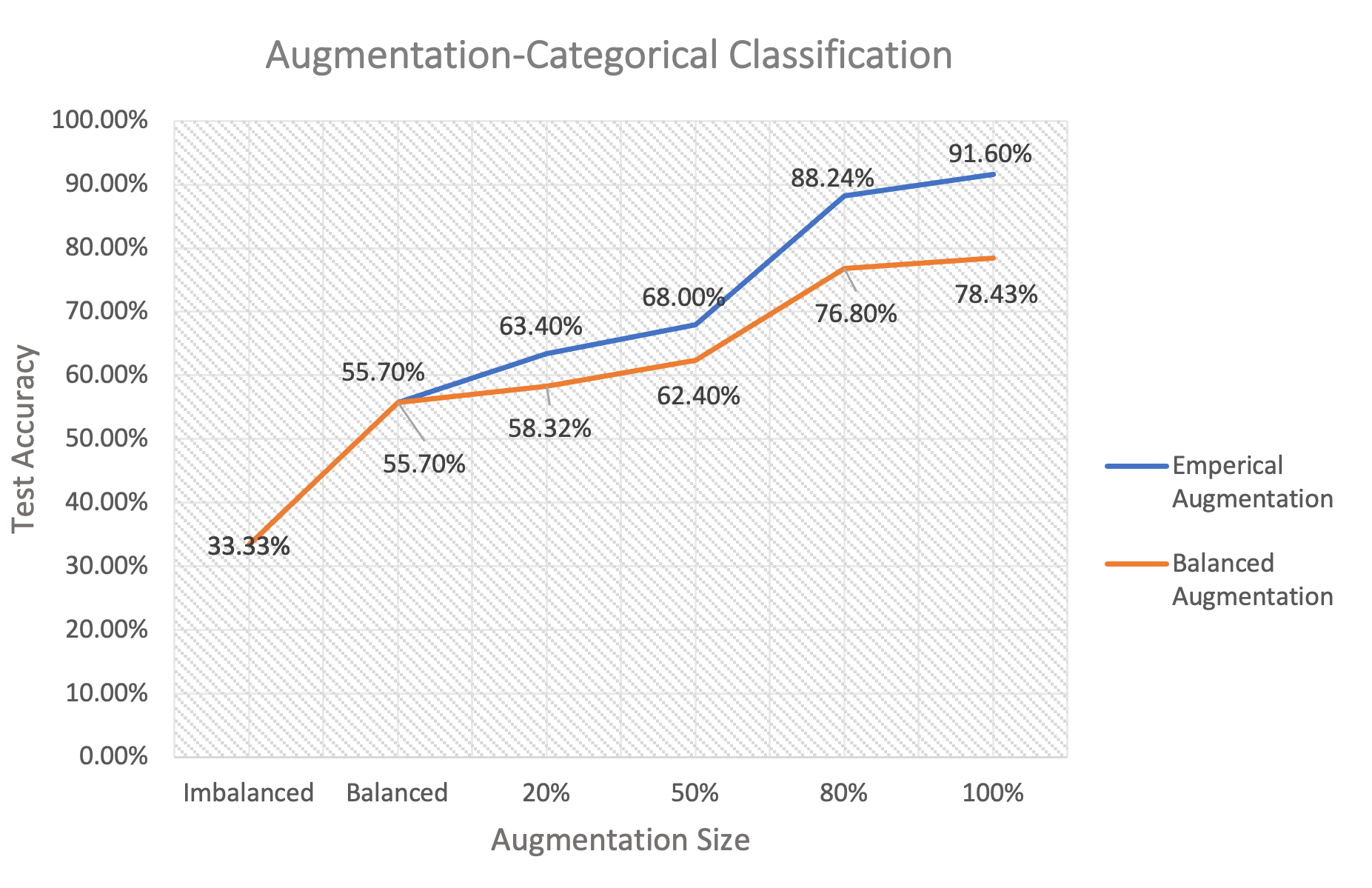} }
\caption{\textbf{Test accuracy of categorical classification for two training cases and a fixed test set}. Case (1): real data that includes only place and speed specific neurons (class 2) and different proportion of balanced simulated data (class 0 and 1). Case (2): real data (class 2), balanced simulated data (class 0 and 1) augmented by the different proportion of empirical simulated data (class 2) while keeping balance.
In both training cases, $100\%$ augmentation size means 4584 neurons.}
\label{fig8}
\end{figure}

Next, we add the animal's speed to the input space and the output includes class 0, 1, and 2. As discussed above, real data contains no example of class 0, also it has no example of class 1 where position signal is exclusively involved. Accordingly, the real data is imbalanced regarding class 0 and 1, and it requires augmentation. We repeat the same procedure as for the binary classification where the test case is fixed, including 50 neurons in class 0 and 50 neurons in class 1 from balanced simulated data, and 50 neurons in class 2 from observed data. For training process, again, we consider two cases: The first case contains 204 neurons from observed data (class 2) that is augmented by various proportion of 4584 neurons from balanced simulated data (1596 neurons class0, 1596 class 1, and 1392 class2), and the second case involves 204 neurons from observed data (class 2) augmented by different fraction of 3192 neurons from balanced simulated data (1596 neurons per class, class 0 and 1), and 1392 neurons from empirical simulated data (class 2).
We illustrates the classification results for these two cases in Fig. \ref{fig8} which indicates how different type of data augmentation can affect test accuracy by balancing as well as enlarging the training size.

Both cases start with using real data alone for training that results in poor test accuracy, $33.33\%$ since the data is imbalance. Then, we add 408 neurons (204 neurons per class) from balanced simulated data for class 0 and 1 to the training set, and test accuracy improves to $55.7\%$. From this point, we increase the training size in two different ways for case one and two. In the first case, we add $20\%$, $50\%$, $80\%$, and $100\%$ of balanced simulated data (including 4584 neurons) respectively to the training set that improves the test accuracy to $58.32\%$, $62.4\%$, $76.8\%$ and $78.43\%$. The test accuracy grows as the training size increasing but it is getting asymptote eventually.
In the second case, we augment the training set by adding $20\%$, $50\%$, $80\%$, and $100\%$ of empirical simulated data (1392 neurons in class 2) while maintaining balance, which leads to $63.4\%$, $68\%$, $88.24\%$ and $91.6\%$ test accuracy respectively. Comparison of these two cases demonstrates that empirical based augmentation leads to higher rate of improvement.

We summarize the classification results based on our statistical model-based data augmentation compared to prior methods without data augmentation in Table \ref{table3}. The results suggest that prior methods cannot achieve better than chance performance for either classification problem; however, the classification results from our proposed data augmentation method are substantially and significantly improved.

\begin{table}
\caption{Comparing classification accuracy between methods without data augmentation  and methods with statistical model-based data augmentation.}
\label{table3}
\setlength{\tabcolsep}{3pt}
\begin{tabular}{|p{90pt}|p{65pt}|p{65pt}|}
\hline
& 
\textbf{Not Augmented}& 
\textbf{Augmented}\\
\hline
Binary Classification& 
$50\%$& 
$98.8\%$\\
\hline
Categorical Classification & 
$33.3\%$& 
$91.6\%$\\
\hline
\end{tabular}
\label{table3}
\end{table}

In the next step, we add direction to the input space, and outputs are class 0 to class 3. Since we have more categories, our ML algorithm requires more data for training. In addition, according to Fig. \ref{fig8}, empirical augmentation for categorical classification indicates higher rate of model performance improvement. Therefore, here we augment the training data including 204 observed neurons (class 3), with 7200 neurons (3600 neurons per class) from balanced simulated data for class 0 and 1, then we add 3600 neurons for class 2 and 3346 neurons for class 3 generated from empirical augmentation. The resulting accuracy was poor although we doubled the training size compare to the previous step. This issue led us to more augmentation, however we had limited computational resources that did not allow for working with a larger dataset. To address this challenge, we makes our ML algorithm (MCDCNN) simpler by changing the number of filters and kernel size of the MCDCNN's layers from 4 and 3 to 3 and 2 respectively. After this change, we obtain $92.3\%$ test accuracy. 

Finally, we build the full classification model by adding the theta phase to the input space, and outputs are class 0 to class 4. In this case, training set contains 204 neurons from the experiment (class 4). First, we augment it with 7200 neurons (3600 neurons per class) from balanced simulated data (class 0 and 1). Then, we add 7200 neurons (3600 neurons per class) for class 2 and 3, and 3346 neurons for class 4 generated from empirical augmentation procedure. Due to limited computational resources, we made the MCDCNN algorithm simpler by changing the number of filters and kernel size from 3 and 2 to 2 and 1 respectively which led to $91.58\%$ test accuracy. The classification results for our methodology are summarized in Table \ref{table5} from the binary classification including class 0 and 1 to the full classification model.

\begin{table}
\caption{Classification results of incremental categories}
\label{table5}
\setlength{\tabcolsep}{3pt}
\begin{tabular}{|p{50pt}|p{30pt}|p{40pt}|p{40pt}|p{50pt}|}
\hline
\textbf{Classes}& 
$\{0, 1\}$& 
$\{0, 1, 2\}$&
$\{0, 1, 2, 3\}$&
$\{0, 1, 2, ,3, 4\}$\\
\hline
\textbf{Training Size} & 
$408$& 
$4788$&
$14350$&
$17950$\\
\hline
\textbf{Test Accuracy} & 
$98.8\%$& 
$91.6\%$&
$92.3\%$&
$91.58\%$\\
\hline
\end{tabular}
\label{table5}
\end{table}

\section{Discussion}
Statistical models provide a powerful approach to study neural encoding at the level of individual neurons or populations ~\cite{paninski2007statistical, truccolo2005point}; however, they are computationally inefficient for analyzing large-scale neural recordings. As experimental methods evolve, more data is recorded over longer periods of time, making it difficult to classify the coding properties of all the neurons in a large population efficiently. Machine learning methods have been shown to provide improved computational efficiency in analyzing large-scale data particularly in neural decoding problems \cite{wen2018neural, horikawa2013neural,glaser2020machine, tampuu2019efficient, saeidi2021neural, livezey2021deep, cruz2014neural,chen2021neural, mathis2020deep}; however, they often require large amounts of training data and labels (for supervised learning), and it can be challenging to assess their quality and interpret the outputs. In this work, we explored the potential for integrating statistical neural encoding models with ML tools for identifying coding properties, and demonstrated how these methods can complement each other to help resolve the limitations of each approach.

This analysis has led to a number of key findings: (1) There are certain classes of network architectures that are specifically well-suited to neural encoding problems since they can identify associations within multivariate time series. Initially, we expected RNN based networks such as LSTMs to be best suited to these datasets \cite{rumelhart1986learning}. LSTMs have been successful in some neural coding analyses, particularly for predicting neural responses and for neural decoding \cite{mcintosh2016deep, batty2016multilayer,glaser2020machine, tampuu2019efficient}. However, they were not able to classify neural population in our analysis, and the potential reason could be their lack of ability to capture the associations between encoded signals and neural responses. CNNs and MCDCNNs were the only network structures in our analysis that were able to classify different neural representations, and MCDCNNs showed substantially faster classification accuracy as a function of training data compared to CNNs.  
(2) Data augmentation can lead to dramatic performance improvement in classification accuracy in cases where there is uneven sampling in the training set. Neuroscientific experiments are often designed so that certain coding properties are undersampled and others are oversampled, and this leads to imbalanced data. This makes the DL classification challenging and augmentation with synthetic data for those coding patterns is a powerful way to address this issue. We used GLMs with prior parameter estimates to generate data for minority classes to balance the training set.  
(3) Not all augmentation schemes are the same. Augmentation with purely theoretical model structures does not provide the same information as augmentation with models that are based on empirical estimates of coding properties from the observed data. The first augmentation scheme is most useful for balancing data, however, once even a small degree of balance is achieved, classification accuracy increases most by augmenting with data that best reflects features in the observed data. Therefore, augmenting data using GLM models with parameters estimated from existing recorded data leads to more powerful classifiers.   

We anticipate our proposed approach could be extremely valuable for large-scale electrophysiology studies of neural coding that are becoming increasingly common \cite{urai2022large, allen2022massive, paninski2018neural, steinmetz2018challenges, vu2018shared}.
For instance, neuropixel analysis has been recently used to study brain-wide systems including auditory, visual, memory, and motor systems using neuropixel probes, for which potentially thousands of neurons might be identified \cite{tasaka2020temporal, barzegaran2022four, deitch2021representational, beshkov2022geodesic, nayebi2022mouse, iqbal2019decoding, ulyanova2019multichannel, park2022motor, sauerbrei2020cortical}. In many of these research areas, recordings are made from a large population of neurons and multiple coding properties that require to be assessed simultaneously \cite{stringer2019spontaneous}. We expect using the extended neural network architectures based on CNNs such as the MCDCNN that we explored here, would be helpful for the reasons demonstrated.

There are a number of limitations to this study that suggest additional analysis to perform. We examined only a small number of ML architectures and selected just one to apply to a particular neural system with a small set of possible covariates. A broader study exploring more architectures, implementing more advanced learning and optimization tools, and applying the methods to broader neural datasets with varied coding properties would provide critical information about the extent to which these data augmentation results generalize. The scope of this study is to to demonstrate that ML methods with specific network architectures can be integrated with statistical models to overcome issues related to limited data and imbalance in the training sets. In our analysis, MLP and LSTM architectures never achieved better than chance performance, however a broader exploration of datasets and estimation methods might suggest situations in which these architectures are successful, or even preferred, for neural classification. In addition, here we focused on coding at the level of individual neurons, but recent trends in neural encoding have focused on interactions between neural spiking and on population level coding. The role of statistical model based data augmentation for those models could be explored by augmenting both the statistical models and ML architectures to include multiple neural responses simultaneously. Moreover, these methods can also be extended to other brain areas, other animals or different types of experiments where the coding variables are unknown.

Statistical methods provide a flexible, interpretable, methodological terminology, and powerful inference methods that have been very successful for understanding neural coding ~\cite{paninski2007statistical, truccolo2005point, pawitan2001all,santner1989statistical,zoltowski2018scaling}. As neuroscience experiments record from larger neural populations over a longer period of time, the need for computationally powerful and efficient models is expanding. On the other hand, ML algorithms demonstrate strong computational power to analyze large-scale data efficiently \cite{luong2014addressing, wu2016google, he2016deep, wen2018neural, horikawa2013neural,glaser2020machine, tampuu2019efficient}. We believe that integrating these two modeling perspectives allows for the development of a statistically principled and computationally efficient paradigm for understanding neural representations. We believe this work is a first step toward addressing the challenges arising from modern neuroscience experiments.

\begin{IEEEbiography}[{\includegraphics[width=1in,height=1.25in,clip,keepaspectratio]{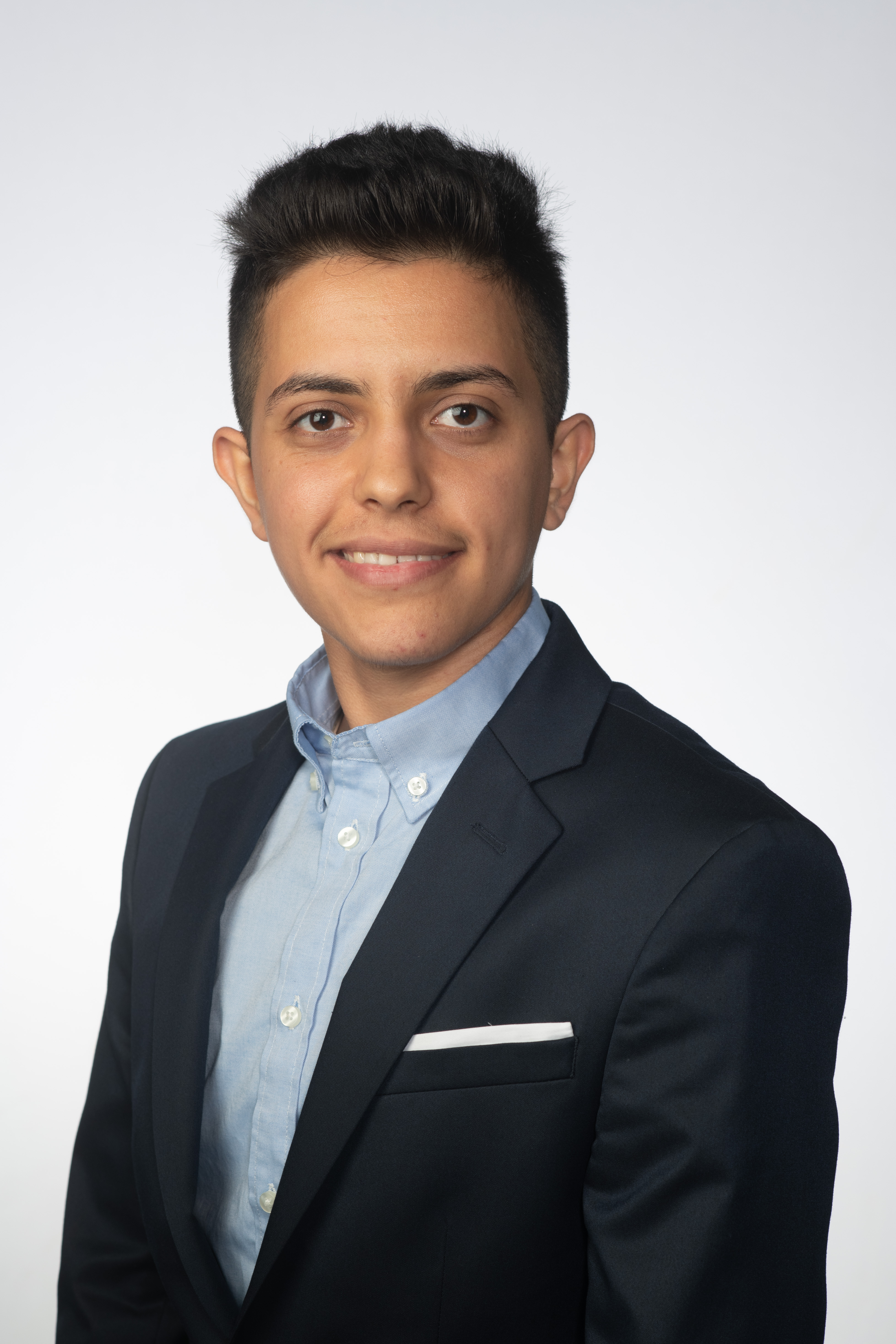}}]{Mehrad Sarmashghi} received the B.S. degree in aerospace engineering from Sharif University of Technology (SUT), Tehran, Iran in 2015. He is currently pursuing the Ph.D. degree in systems engineering at Boston University, Boston, MA, USA.

From 2016 to 2022, he was a Research Assistant with the Division of Systems Engineering at Boston University, Boston, MA. His research interest includes the development of statistical model identification methods to analyze neural spiking data, and integrating them with machine learning algorithms to classify neural populations. 

\end{IEEEbiography}

\begin{IEEEbiography}[{\includegraphics[width=1in,height=1.25in,clip,keepaspectratio]{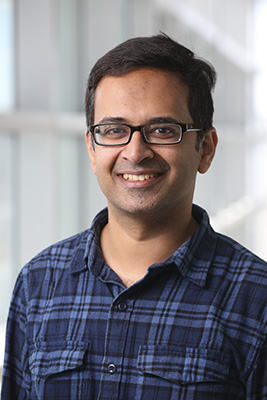}}]{Shantanu P. Jadhav} received the B.S. degree in engineering physics from Indian Institute of Technology (IIT), Bombay, India, in 2002 and the Ph.D. degree in biology from University of California San Diego (UCSD), San Diego, California, USA, in 2008. He has been a Postdoctoral Fellow in neuroscience at University of California San Francisco (UCSF) and University of California Berkeley (UCB) from 2008 to 2014.

From 2002 to 2003, he was a Research Assistant in Neuroscience at National Center for Biological Sciences (NCBS), India. Since 2014, he has been an Associate Professor with the Department of Psychology at Brandeis University, Waltham, Massachusetts, USA. His research interest is focused on understanding the neural basis of cognitive abilities by studying processing at the cellular and network level in the neuronal circuits of the rodent brain.

\end{IEEEbiography}

\begin{IEEEbiography}[{\includegraphics[width=1in,height=1.25in,clip,keepaspectratio]{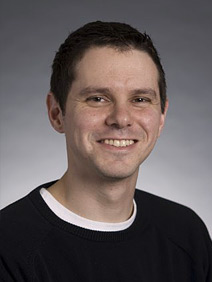}}]{Uri T. Eden} received the B.S. degree in mathematics and in engineering and applied sciences from the California Institute of Technology, Pasadena, California, USA in 1999, and the S.M. degree in engineering sciences and the Ph.D. degree in engineering sciences from the Division of Engineering and Applied Sciences at Harvard University, Cambridge, MA, in 2002 and 2005. From 2005-2006, he was a Postdoctoral Fellow in Statistical Neural Data Analysis with the Neuroscience Statistics Research Laboratory at Massachusetts General Hospital (MGH), Charlestown, MA. 

In 2006, he joined the Department of Mathematics and Statistics at Boston University, Boston, MA, where he is currently an Associate Professor and the Assistant Director of the Program in Statistics. He received an NSF CAREER award in 2007. His research focuses on developing mathematical and statistical methods to analyze neural spiking activity, integrating methodologies related to model identification, statistical inference, signal processing, and stochastic estimation and control.

\end{IEEEbiography}

\EOD

\end{document}